\documentstyle[12pt]{article}
\newcommand{\beq}{\begin{equation}}
\newcommand{\eeq}{\end{equation}}
\newcommand{\beqa}{\begin{eqnarray}}
\newcommand{\eeqa}{\end{eqnarray}}
\newcommand{\bega}{\begin{array}}
\newcommand{\enar}{\end{array}}
\newcommand{\beqs}{\begin{displaymath}}
\newcommand{\eeqs}{\end{displaymath}}
\newcommand{\beqas}{\begin{eqnarray*}}
\newcommand{\eeqas}{\end{eqnarray*}}
\newcommand{\begas}{\begin{eqnarray*}}
\newcommand{\enars}{\end{eqnarray*}}
\newcommand{\dslash}{\not\hspace{-1mm}\nabla}
\newcommand{\dslashf}{\not\nabla}

\newcommand{\Dslash}{/\hspace{-3mm}}
\newcommand{\bd}{\partial}
\newcommand{\befi}{\begin{figure}}
\newcommand{\efi}{\end{figure}}
\begin{document}
\title{Casimir Driven Evolution of the Universe}
\author{Frank Antonsen \& Karsten Bormann\\Niels Bohr Institute\\
Blegdamsvej 17\\ 2100 Copenhagen \O \\ Denmark}
\maketitle
\begin{abstract}
For a Friedman-Robertson-Walker space-time in which the only
contribution to the stress-energy tensor comes from
the renormalised zero-point energy (i.e. the Casimir energy) of the fundamental
fields the evolution of the universe (the scale factor)
depends upon whether the universe is open, flat or
closed and upon which fundamental fields inhabit the
space-time. \\
We calculate this ``Casimir effect'' using the heat kernel 
method, and the calculation is thus non-perturbative. We treat
fields of spin $0,\frac{1}{2},1$ coupled to the gravitational
background only. 
The heat kernels and/or zeta-functions for the various spins are
related to that of a non-minimally coupled scalar field which is
again related to that of the minimally coupled one. A WKB approximation
is used in obtaining the radial part of that heat kernel.\\
The simulations of the resulting equations of motion seem to exclude the 
possibility of a closed universe, $K=+1$,
as these turn out to have an overwhelming tendency towards a fast collapse -- the details
such as the rate of this collapse depends on the structure of the underlying 
quantum degrees of freedom; a non-minimal coupling to curvature accelerates the
process. Only $K=-1$ and $K=0$ will in general lead to macroscopic
universes, and of these $K=-1$ seems to be more favourable.\\
The possibility of the scale factor
being a concave rather than a convex function, as is the case for a $K=-1$ FRW-spacetime
inhabited by a conformally coupled scalar field, potentially indicates that the problem
of the large Hubble constant is non-existent as the age of the universe need not
be less than or equal to the Hubble time. \\
Note should be given to the fact, however, that we are not able to pursue 
the numerical study to really large times neither to do simulations for a
full standard model.
\end{abstract}

\section{Introduction}
In a recent paper \cite{prd}, we realized that the Casimir effect
may drive inflation or otherwise strongly influence the evolution
of space-time because the vacuum fluctuations
couple to the (background) gravitational field. As one cannot escape the
Casimir effect it may thus be that this property of the
zero-point energy can be used for determining whether
the universe is open, flat or closed. Or working the other way around,
one may be able to use cosmological observations to tell us about the
particle content in the very early universe. \\
In this paper the quantum fields reside in a 
Friedman-Robertson-Walker space-time. In section 2 we show how one
in general calculates
the Casimir effect of a free scalar field using the heat kernel
method and continue to determine it explicitly in section 3 in the case of the
Friedman-Robertson-Walker metric and it is noted that using
this approach, the zero-point energy is almost automatically renormalised.
The renormalised Casimir energy is then inserted in section 3.1 into the
right hand side of the Einstein equations (it gives an effective action for
the scale factor $a(t)$, the only degree of freedom)
in order to obtain the equations of motion and the results
of the corresponding simulations of the evolution of the space-time,
perhaps relevant if the Higgs field is
believed to dominate the early universe, are presented in section 3.2. 
Having established a procedure for determining the Casimir effect on
the evolution of space-time we continue to consider the case of
a non-minimally coupled scalar field (in section 4) and show how the
Casimir energy is related to that of the minimally coupled scalar
field. Then we proceed in sections 4.1 and 4.2 to
present the equations of motion and the corresponding simulations
for a conformally coupled scalar field. In section 5 the approach
used in section 4 is generalised to the case of vector bosons, with the 
problematic
self-interaction being treated using a mean field approximation
procedure - due to the complexity of the solution we have
not been able to carry out the simulations, however. Finally in section 6,
after some general considerations
about how to link a fermionic operator to a scalar or bosonic one,
we take on free (spin $1/2$) fermions. For completeness,
we have included a discussion of fermions interacting with gauge fields
in an appendix (based on the mean field
approach developed for pure Yang-Mills theory).
Einstein equations and simulations for free fermions are presented in
sections 6.1 and 6.2
Section 7 gives a few concluding remarks.

\section{General Considerations concerning the calculation of
 the Casimir Effect for a Free Scalar Field}
We need to determine the right hand side of the Einstein equations,
the energy-momentum tensor. Now, the quantum expectation value, $\langle T_{\mu\nu}
\rangle$, (the zero-point energy) is given in terms of the effective 
action $\Gamma$ as \cite{BD,GMM}
\beq
    \langle T_{\mu\nu}\rangle = 2(-g)^{-1/2}\frac{\delta \Gamma}{\delta g^{\mu\nu}}
\eeq
which is just the quantum analogue of the classical relationship
\beqs
    T_{\mu\nu}^{\rm class} = -g^{-1/2}\frac{\delta S}{\delta g^{\mu\nu}}
\eeqs
Another way of looking at this (the way chosen in \cite{prd}), is to use an analogy
with classical statistical mechanics and introduce 
Helmholtz' free energy, $F$, which is related to the partition function by
\begin{equation}
    F = -\frac{1}{\beta}\ln Z
\end{equation}
where $\beta$ is the inverse temperature
and where the partition function is given by the functional integral
\begin{equation}
    Z = \int e^{-S}{\cal D}\phi = e^{-\Gamma_{\rm eff}}
\end{equation}
with $\Gamma_{\rm eff}$ being the effective action, i.e.
\beq
    \Gamma = \beta F = -\ln Z
\eeq
Actually, this is an effective potential for the gravitational degrees of freedom, the full
effective action for gravity being given by
\beq
    \Gamma_{\rm full}[g] = S_{\rm EH}[g]+\Gamma[g]
\eeq
but we will always take the Einstein-Hilbert action $S_{\rm EH}$ to be understood, and 
simply refer to $\Gamma$ as the effective action. This slight abuse of notation should not
cause any misunderstandings.\\
>From Helmholtz' free energy one can then, as in \cite{prd}, proceed to 
find the internal energy and the pressures. It is because of this 
analogy with the free energy we refer
to the entire scenario as Casimir driven evolution, the renormalised zero point
energy being analogous to the well known Casimir effect in flat space.\\
Since the effective action is given by integrating out the matter degrees of freedom
in the partition function, it gives the effective equation of motion for the
remaining degree of freedom, namely the gravitational one, which for a Friedman-Robertson-Walker
background reduces to an effective equation of motion for the scale factor $a(t)$.\\
Now, since we're dealing with a free particle,
\begin{equation} 
      S=\frac{1}{2}\int d^4x \sqrt{-g}(g^{\mu_\nu}\partial_\mu 
      \phi\partial_\nu \phi-M^2\phi^2)
\end{equation}
 albeit in a special
geometry, this integral is merely a Gaussian. Thus we can perform it using
standard techniques, arriving at the result
\begin{equation}
     Z = (\det A)^{-1/2}    \label{eq:det}
\end{equation}
where the differential operator $A$ is given by
\begin{equation}
    A \equiv \frac{\delta^2 S}{\delta \varphi^2} = \Box-M^2
\end{equation}
with $\Box$, the d'Alembertian of the curved space-time, given by \cite{BD}
\beq
\Box=\frac{1}{\sqrt{|g|}}\partial_\mu\left(g^{\mu\nu}\sqrt{|g|}
      \partial_\nu\right)  \label{eq:def-box}
\eeq
\\
The effective action is thus seen to be given by a functional determinant;
\begin{equation}
    \Gamma=-\ln \left(\det \left(\Box-M^2\right)\right)^{-1/2} \label{eq:determinant}
\end{equation}
The determinant of an operator, $A$, is, by definition, 
the product of its eigenvalues
\begin{equation}
\det(A)=\prod_\lambda \lambda
\end{equation}
The zeta-function is given by \cite{zeta,Ramond}
\begin{equation}
\zeta_A(s)=\sum_\lambda \lambda^{-s} = {\rm Tr}~A^{-s}
\end{equation}
and it is related to $det(A)$ by the following equation
\begin{equation}
	\left.\frac{d\zeta_A}{ds}\right|_{s=0}
		=\sum_{\lambda}-\ln\lambda\cdot\lambda^{-s}|_{s=0}
                          =-\sum_\lambda \ln\lambda
                          =-\ln\prod_\lambda\lambda=-\ln\det(A)
\end{equation}
Now (as shown below) the zeta-function, $\zeta_A$, can be determined from  
the heat kernel $G_A(x,x',\sigma)$  which is a
function satisfying the heat kernel equation\footnote{If
$A$ is the Laplace operator and the space is $S^1$ then $\zeta(s)$ becomes 
(proportional to) the Riemann zeta function,
and if the space is $R^3$ and $A=\nabla^2$ the heat kernel equation becomes
the usual heat equation.}
\begin{equation}
    A_x G_A(x,x',\sigma) = -\frac{\partial}{\partial\sigma}G_A(x,x',\sigma)
\end{equation}
subject to the boundary condition $G_A(x,x',0) =\delta(x,x')$.\\
To see how $G_A(x,x',\sigma)$ is related to $\zeta_A$ note 
that if $A\phi_\lambda=\lambda\phi_\lambda$ then 
$G_A(x,x',\sigma)\equiv\sum_\lambda\phi_\lambda(x)\phi_\lambda^*(x')
e^{-\lambda\cdot\sigma}$ satisfies the heat kernel equation. This
in turn makes it possible to show the following relationship 
between $G_A$ and $\zeta_A$
\begin{eqnarray*}
      \int_0^\infty d\sigma \sigma^{s-1}\int \sqrt{g}d^4x G_A(x,x,\sigma)
      &=&\int_0^\infty d\sigma \sigma^{s-1}\int \sqrt{g}d^4x\sum_\lambda|
      \phi_\lambda(x)|^2e^{-\lambda\sigma} \\
      &=&\sum_\lambda\int_0^\infty d\sigma \sigma^{s-1}e^{-\lambda\sigma} 
                 \int \sqrt{g}d^4x|\phi_\lambda(x)|^2
\end{eqnarray*}
where, with suitable normalisation, the last integral can be put equal
to unity
%\footnote{One shouldn't think of $\phi_\lambda$ as a wave function 
%(the integral might then be un-bounded), rather think of it as the 
%eigenvector of a (hermitian) operator  -- these can always be chosen to be orthonormal.} 
which gives us
\begin{eqnarray}
\int_0^\infty d\sigma \sigma^{s-1}\int \sqrt{g}d^4x G_A(x,x,\sigma)
&=&\sum_\lambda\int_0^\infty d\sigma \sigma^{s-1}e^{-\lambda\sigma}\nonumber\\
&=&\sum_\lambda \lambda^{-s}\int_0^\infty d(\lambda\sigma) (\lambda\sigma)^{s-1}
                                              e^{-\lambda\sigma}\nonumber\\
&=&(\sum_\lambda \lambda^{-s})\Gamma(s)\nonumber\\
&=&\Gamma(s)\zeta_A(s)
\end{eqnarray}
so that
\begin{displaymath}
    \zeta_A(s) \equiv \frac{1}{\Gamma(s)}\int d\sigma \sigma^{s-1}\int G_A(x,x,\sigma)\sqrt{g}d^4x
\end{displaymath}
and so, finally, for the minimally coupled scalar case
\beq
      \Gamma =\frac{1}{2}\int \sqrt{g}d^4x\left.\frac{d}{ds}
      \right|_{s=0}\int^\infty_0 d\sigma
\frac{\sigma^{s-1}}{\Gamma(s)}G_{\Box-m^2}(x,x,\sigma)\label{eq:f1}
\eeq
Note that the integral over $x$ is taken along the diagonal $x=x'$.  
Also note that the $\sigma$-integral
is a standard one if one uses the spectral representation of the 
zeta-function, 
thus easy to calculate though it generally needs further regularisation (in
the case of a minimally coupled field we use the spectral representation of
the heat kernel (in a WKB approximation) while in other cases we use a hybrid 
expression, partially
referring to eigenfunctions, partly a Taylor series generalising the usual
Schwinger-DeWitt expansion \cite{SDW}), including the exponential of the
scalar curvature as shown by Jack and Parker, \cite{JP}, to be needed.
\\
Thus to calculate the zero-point energy of a scalar field
one determines the scalar field operator (the d'Alembertian)
in the relevant space-time, solve the corresponding heat
kernel equation and subsequently calculates the zeta-function
from which the generating functional and thus all relevant
quantities (including the zero-point energy) can easily be
calculated.\\
The zeta-function of spin 1 gauge bosons can be determined by
a generalisation of the procedure for the non-minimal scalar field
while
the zeta-function associated with 
fermions is directly related to that of the
a non-minimally coupled scalar field. This in turn is related to the 
zeta-function of a minimally coupled scalar field, 
substantially reducing the amount of work involved in determining
the zero-point energy of these fields, as will be shown in the following 
sections.

\section{Solving the Heat Kernel Equation for the Minimally Coupled
 Scalar Field}
Having in principle established a procedure for determining the zero-point
energy for a minimally coupled scalar field the next step is to find 
the heat kernel and subsequently the Casimir energy of the field explicitly.
The heat kernels for higher spin, as well as for non-minimally coupled
scalar fields, are going to be related to the result found in this section. 
Thus this calculation is a central part of the paper.\\
If one writes the Friedman-Robertson-Walker line-element with the scale factor 
$a=a(t)$ as
\beq
    ds^2=dt^2-a^2(t)(d\chi^2+f^2(\chi)(d\theta^2+\sin^2(\theta)d\phi^2))
	\label{eq:metric}
\eeq
where 
\beq
f(\chi)=\left\{\bega{ll} \sin\chi&\mbox{for $K=+1$  (closed universe)}\\
                       \chi        &\mbox{for $K=0$   (flat universe)}\\
                       \sinh\chi &\mbox{for $K=-1$  (open universe)} 
      \enar\right.
\eeq
then the d'Alembertian becomes
\beqa
\Box &=& \frac{1}{\sqrt{|g|}}\partial_\mu\left(g^{\mu\nu}\sqrt{|g|}
      \partial_\nu\right)\\
     &=& (\partial_t^2+3\frac{\dot{a}}{a}\partial_t)
         -\frac{1}{a^2}\left(\bd_\chi^2-2\bd_\chi\left(\frac{\bd_\chi f}{f}
      \bd_\chi\right)
    \right)-\frac{L^2}{a^2f^2}\label{eq:dalem}
\eeqa
where $L$ is the usual angular momentum operator.\\
Even though the equation separates, the minimally coupled scalar heat kernel 
equation
\beqa
    \Box G_{\Box}(x,x',\sigma)  &=&
    \left((\partial_t^2+3\frac{\dot{a}}{a}\partial_t)
    -\frac{1}{a^2}\left(\bd_\chi^2-2\bd_\chi(\frac{\bd_\chi f}{f}\bd_\chi)
      \right)-
    \frac{L^2}{a^2f^2}\right)G_{\Box}(x,x',\sigma)\nonumber\\ 
    &=&-\frac{\partial}{\partial \sigma}G_{\Box}(x,x',\sigma) 
\eeqa
is rather hard to solve. Fortunately, the eigenfunctions,  $\psi_\lambda({\bf 
x})$, of the spatial part of the d'Alembertian are already known to be 
\cite{BD,FRWsol}
\beq
\psi_\lambda({\bf x})=\left\{\bega{ll} 
         (2\pi)^{-\frac{3}{2}}e^{i{\bf k}\cdot{\bf x}}
             \mbox{    where  }{\bf k}=(k_1,k_2,k_3) &\mbox{    for }K=0\\
       \Pi^{(\pm)}_{kJ}(\chi)Y^M_J(\theta,\phi) 
            \mbox{  where  } k=(k,J,M)  &\mbox{    for }K=\pm 1 \enar\right.
\eeq
where
\beqa
   && -\infty < k_i < \infty\mbox{  ; where   } k=|{\bf k}|\mbox{   for $K=0$}
      \nonumber \\
   && M=-J, -J+1, ... , J ;\left\{\bega{ll} J=0, 1, ... , k-1 &; \mbox{  } 
      k= 1, 2, ... \mbox{    for }K=1\\
          J=0, 1, ... &; \mbox{  } 0< k< \infty  \mbox{    for }K=-1\enar\right.
\eeqa
and
\beqa
    \Pi^{(-)}_{kJ}(\chi)&=&
    \left\{\frac{1}{2}\pi k^2(k^2+1)...(k^2+(2+J)^2)\right\}
    ^{-\frac{1}{2}}\sinh\chi
    \left(\frac{d}{d\cosh\chi}\right)^{1+J}\cos(k\chi)
\nonumber\\
    \Pi^{(+)}_{kJ}(\chi)&=&\left\{\frac{1}{2}\pi (-k^2)(-k^2+1)...(-k^2+
      (2+J)^2)\right\}^{-\frac{1}{2}}\nonumber\\
    &&\hspace{50mm}\times i\sin\chi\left(\frac{d}{d\cos\chi}\right)^{1+J}\cos(k\chi)
\eeqa
as quoted by \cite{BD} so proceed by assuming the heat kernel of the
d'Alembertian to be of the form
\beq
 G_\Box=\sum_\lambda \psi_{\lambda}({\bf x})\psi_{\lambda}^*({\bf x'})
    T_\lambda(t,t',\sigma)
\eeq
which is a solution of the heat kernel equation provided that the set
of functions $T_\lambda(t,t',\sigma)$
solves 
\beq
    (\partial_t^2+3\frac{\dot{a}}{a}\partial_t)T_\lambda (t,t',\sigma)=
    -\bd_\sigma T_\lambda (t,t',\sigma)
\eeq
Writing
\beq
    T_\lambda(t,t',\sigma) = a^{-3/2}(t)f_\lambda(t)a^{-3/2}(t')f_\lambda(t')
    e^{-\lambda\sigma}
\eeq
gives the following equation for the function $f_\lambda$
\beq
    \ddot{f_\lambda}=\left(\lambda+\frac{27}{4}\left(\frac{\dot{a}}{a}\right)^2-
    \frac{3}{2}\frac{\ddot{a}}{a}-\lambda a^{-2}
    \right)f_\lambda\equiv h_\lambda(t)f_\lambda(t)
\eeq
with
\beq
    \lambda =k^2-K
\eeq
being the eigenvalue of the spatial part of the d'Alembertian.
\\
We have not been able to solve this equation explicitly (due to the fact that 
we have no {\em a priori} knowledge about $h_\lambda(t)$). Note however the 
similarity with the stationary Schr\"{o}dinger equation; $\frac{2m}{\hbar}
[E-V(x)]\psi(x)=\triangle\psi(x)$. Thus if $\frac{\dot{h}}{\sqrt{h}}$ 
is negligible it is sound to use the WKB-approximation\footnote{This 
approximation assumes that $h$ and thus the scale factor $a$ is relatively
stable and, in turn, tends to stabilise $a$. Thus, when simulating the 
behaviour of the scale factor later on, the evolution of $a$ will probably 
be understated.} giving 
\beq
    f_\lambda = h^{-\frac{1}{4}}_\lambda(t) e^{\int_0^t dt' 
      \sqrt{h_\lambda}(t')}
\eeq
One should note that this is the only approximation introduced so far.
\\
Having obtained a set of functions on which to expand the heat kernel; 
\beqa
G_\Box(x,x',\sigma)&=&
\sum_\lambda\phi_\lambda(x)\phi_\lambda^*(x')
e^{-\lambda\cdot\sigma}\nonumber\\
&=&\sum_\lambda\psi_{\lambda}({\bf x})\psi_{\lambda}^*({\bf x'})
    T_\lambda(t,t',\sigma)\nonumber\\
&=&\sum_\lambda\psi_{\lambda}({\bf x})\psi_{\lambda}^*({\bf x'})
a^{-3/2}(t)f_\lambda(t)a^{-3/2}(t')f_\lambda^*(t')e^{-\lambda\sigma}\nonumber\\
&=&\sum_\lambda\psi_{\lambda}({\bf x})\psi_{\lambda}^*({\bf x'})
a^{-3/2}(t)h^{-\frac{1}{4}}_\lambda(t) e^{\int_0^t dt''
      \sqrt{h_\lambda}(t'')} \times\nonumber\\
&&\hspace{30mm}a^{-3/2}(t')h^{-\frac{1}{4}}_\lambda(t') 
e^{\int_0^{t'} dt''' 
      \sqrt{h_\lambda}(t''')} e^{-\lambda\sigma}\label{eq:heat0}
\eeqa
the $\zeta$-function reads
\beqa
    \zeta_\Box(s) &=& \frac{1}{\Gamma(s)}\sum_\lambda \int_0^\infty \int
    |\psi_\lambda({\bf x})|^2h^{-1/2}(t)e^{-2\int_0^t\sqrt{h_\lambda(t')}dt'}
    f^2(\chi)\sqrt{g} e^{\lambda\sigma}\sigma^{s-1}d^3x dtd\sigma\nonumber\\
    &=& \frac{1}{\Gamma(s)}\sum_\lambda\int_0^\infty\int h^{-1/2}(t)
    e^{-2\int_0^t\sqrt{h_\lambda(t')}dt'}e^{-\lambda\sigma}\sigma^{s-1}dtd\sigma
    \label{eq:zeta}
\eeqa
where we have used $\sqrt{g} = a^3f^2\sin\theta \equiv a^3\sqrt{g^{(3)}}$ and 
that the eigenfunctions of the spatial part of the d'Alembertian are 
orthonormal with respect to the measure $\sqrt{g^{(3)}}d^3x = f^2(\chi)d\chi 
d\Omega$. 

\subsection{The Einstein Equations for the Friedman-Robertson-Walker space-time
with a minimally coupled scalar field}
With the metric given by equation (\ref{eq:metric}) 
the non-zero components of the Einstein 
tensor are ($f'=\partial_\chi f$)
\beqa
    G_{00}=-\frac{2ff''+f'^2-3\dot{a}^2f^2-1}{a^2f^2}\nonumber\\
    G_{11}=-\frac{f'^2+(-2a\ddot{a}-\dot{a}^2)f^2-1}{a^2f^2}\nonumber\\
    G_{22}=-\frac{f''+(-2a\ddot{a}-\dot{a}^2)f}{a^2f}\nonumber\\
    G_{33}=-\frac{f''+(-2a\ddot{a}-\dot{a}^2)f}{a^2f}
\eeqa
Inserting $f$ from equation (19) it is explicitly seen that, incidentally, 
the (two) last components can be rewritten as
\beq
    G_{33}=G_{22}=\frac{2a\ddot{a}+\dot{a}^2+K}{a^2}
\eeq
leading to the following (fourth component of the) equation of motion for the
scale factor $a(t)$ (using the first or second component would have complicated
things unnecessarily as $t$ would no longer be the only variable)
\beq
    \frac{2a\ddot{a}+\dot{a}^2+K}{a^2}= \langle T_{33}\rangle
	\label{eq:eom}
\eeq
One should note that while this may sound
a bit arbitrary (involving a choice of using only $G_{33}$), it is actually the uniquely
defined equation of motion which follows from the effective action. The full Einstein
tensor is only listed in order to give an interpretation of the resulting equation, namely
as the pressure coupled to $G_{33}$.\\ 
We proceed to calculate the resulting effective action
explicitly using equations (\ref{eq:f1},\ref{eq:heat0},\ref{eq:zeta}):
\beqa
    \Gamma&=&\frac{1}{2}\left.\frac{d\zeta_A}{ds}\right|_{s=0}\\
    &=&\left.\frac{d}{ds}\right|_{s=0}\frac{1}{\Gamma(s)}\lambda^{-s}
       \sum_\lambda\Gamma(s)\int h_\lambda(t)^{-1/2}e^{-2\int_0^t\sqrt{h_\lambda}
       dt'}dt\nonumber\\
    &=& -\frac{1}{2}\sum_\lambda\ln\lambda\int h_\lambda(t)^{-1/2}
       e^{-2\int_0^t\sqrt{h_\lambda}dt'}dt
\eeqa
with
\beq
    h_\lambda = 3\dot{a}a^{-2}-\frac{3}{2}\ddot{a}a^{-1}+\lambda+a^{-2}\lambda
\eeq
coming from the WKB-approximation. The right hand side of the
equation of motion (\ref{eq:eom}) for the scale factor $a(t)$, obtained by varying $\Gamma$
with respect to $a(t)$, thus reads
\beqa
    &&\hspace{-30mm}-\sum_\lambda\ln\lambda\left\{h_\lambda^{-3/2}
      \left(-\frac{3}{2}\dot{a}
    a^{-3}+\frac{3}{4}\ddot{a}a^{-2}-\lambda a^{-3}\right)+\right.
    \nonumber\\
    &&\hspace{10mm}\left. h_\lambda^{-1/2}\int_0^t\frac{-6\dot{a}a^{-3}+
      \frac{3}{2}\ddot{a}
    a^{-2}-2\lambda a^{-3}}{\sqrt{3\dot{a}a^{-2}-\frac{3}{2}\ddot{a}a^{-1}+ 
    \lambda+a^{-2}\lambda}}dt'\right\}e^{-2\int_0^t\sqrt{h_\lambda}dt'}
\eeqa
We note the appearance of an integral over time: Formally the system depends on
its entire past. Even though this dependence makes simulations
somewhat tedious, we have still carried out quite a few, and these will be
discussed in the next subsection.\\
These equations differ from the ones found by Suen and Anderson, \cite{SA},
due to the use of the WKB-approximation. Note by the way that it is the
exponential coming from this approximation which ensures rapid convergence of
the summation over $\lambda$, thereby making the numerical solution more
feasible.

\subsection{Simulating the Equations of Motion}
In order to simulate the evolution of a Friedman-Robertson-Walker spacetime 
according to
the above equation of motion for $a(t)$, we first have to discretise
the time, i.e., to introduce a time-step $\delta t$. One immediate problem 
is the
appearance on the right hand side (the free energy contribution) of $a(t)$ and
its derivatives raised to various powers and exponentiated. Thus we cannot just
insert $\dot{a}(t) = (a(t+\delta t)-a(t))/\delta t$, $\ddot{a}(t) = 
(a(t+\delta t)- 2a(t)+a(t-\delta t))/\delta t^2$ into the equation of 
motion and
then find a recursion relation telling us how to compute $a(t+\delta t)$ from 
the
knowledge of $a(t),a(t-\delta t)$, as we are not able to isolate $a(t+\delta t)
$. Instead we will consider the right hand side as a source term to be 
evaluated
at $t-\delta t$ whereas the left hand side, the Einstein tensor, is to be
evaluated at time $t$. One can either justify this by sheer necessity or 
argue that it is plausible that the disturbance caused by the vacuum
fluctuations do not give rise to an instantaneous change of geometry. In the
limit $\delta t\rightarrow 0$ this distinction, of course, disappears. With
this, we can now isolate $a(t+\delta t)$ as it only appears in the discretised
version of the Einstein tensor.\\
As mentioned in the introduction this work was inspired by the simulations
of the (space-time called a) hyper-spatial tube. Those simulations did show
a number of qualitatively different types of behaviour, the behaviour being
determined by the initial conditions. Thus we have varied the initial 
conditions which for $a(t=0)$, in units of the Planck length, were
$3,5$ or $10$ and  for $\dot{a}(t=0)$, in units of the velocity of light, were
$0.0, \pm 0.1, \pm 0.5 \pm 1.0$ and $1.5$\footnote{These are not observable 
velocities
but geometrical ones and thus potentially can be larger than $c$.}.

Figures 1a-c shows, for the various initial conditions, simulations for 
$K=+1,0$ 
and $-1$ respectively. As can be seen, the behaviour of the scale factor is
rather universal in the sense that for given $K$ the initial conditions 
determine whether the universe will collapse or expand but once it expands 
there is little
(qualitative) dependence on the initial conditions of the evolution.
\\
We notice that for $K=+1$ there is a marked tendency to collapse, whereas for
the two other cases $K=-1,0$ we have a tendency to expansion. 
Due to the amount of computer time and storage required, we have 
unfortunately not been able to pursue these evolutions very far, and so
extrapolations on the evolution for large $t$ becomes somewhat speculative. 
However, to see if the behaviour was qualitatively stable at somewhat higher
times we prepared a plot of the evolution for $K=-1$ up to times 
$10^3 T_{Planck}$, this did not give rise to any new behaviour
and is therefore not included here. 
Also note that the expansion becomes polynomial at late 
times\footnote{This is not obvious from 
the plots, but by ignoring the data from first $10 T_{Planck}$ one gets data
that indeed are on straight lines in a doubly logarithmic plot.}  
(powers $0.6-0.7 $ for $K=0$ and $\sim 0.8$ for $K=-1$). 
\\
The early paper by
Fischetti, Hartle and Hu \cite{FHH} found a solution for $K=0$ to be
of the assymptotic form $a(t)\sim \sqrt{t}$, which was corrected by
Starobinsky \cite{Star} a few years later where he found $a(t)\sim a_1 
t^{2/3}(1+(2/3Mt)\sin M(t-t_1))$, which is consistent with our findings. 
\\
It
should be noted, however, that both of these papers had to introduce unknown
constants, $k_2,k_3$ (in the notation of Starobinsky), which they then
supposed to be both positive (Fischetti et al.) or $k_2>0, k_3<0$ 
(Starobinsky). These quantities should be determined from the summation over 
quantum degrees of freedom. Our approach has no such arbitrary constants since
we start from first principles, and hence these constants take on definite
values which apparently are consistent with Starobinsky's results. 
\\
Moreover,
both of these early papers take the energy momentum tensor to obey $p=
\frac{1}{3}\rho$, which is not necessarily true for quantum corrections, and
moreover do not violate any energy conditions as the true quantum field
theoretical vacuum supposedly does. Since we use the Casimir energy density
our energy momentum tensor generally violates the classical energy conditions, as
one would expect a quantum vacuum to do.
\\
The FRW-space-time is somewhat similar to that of the hyperspatial tube 
\cite{prd} (which is
a cylinder in 5d, i.e in 4d it has the topology $R\times R\times S^2$). 
Therefore the universal behaviour as well as the slow expansion rate is 
rather surprising: In analogy with the case of the hyperspatial tube
 (in one of the five qualitatively different scenarios), we 
get expansion by first decreasing the size of the universe a bit; at some 
point, then, a repulsive force sets in, and the expansion starts. 
Looking at the data in a double logarithmic plot shows this clearly. 
But in the case of the hyperspatial tube 
this repulsive force ultimately gave rise to inflation-like
expansions followed by polynomial growth (cf figure 5b of \cite{prd}), 
this is not seen here. 
\\
It is of course still possible that such a type of expansion can occur and that
we simply missed it due to our choice of initial conditions\footnote{The 
calculations done in this paper are `fully dynamical' while those of 
\cite{prd} are quasi-static. However, we would not expect that this is what 
causes the difference.} (or that some other residing field(s) than those 
considered in this paper could turn on such behaviour).

\section{The Non-Minimally Coupled Case}
The heat kernel equation for a non-minimally coupled scalar field is
\beq
    (\Box+\xi R)G_{\Box+\xi R}(x,x',\sigma)=-\bd_\sigma G_{\Box+\xi R}
    (x,x',\sigma) \label{eq:nonminG}
\eeq
where $R= 6[a\ddot{a}+\dot{a}^2+K]/a^2$  ($K=0, \pm 1$) is the curvature 
scalar. We know how to solve
this for $\xi=0$ (i.e. minimal coupling), and the solution to the generally 
coupled heat kernel equation is now assumed to 
be of the form 
\beq
    G_{\Box+\xi R}(x,x',\sigma)=G_\Box(x,x',\sigma)F_\xi 
      (x,x',\sigma)
\eeq
where $G_\Box$ is the heat kernel for the minimally coupled case ($\xi=0$).
Inserted into the heat kernel equation this gives
\beq
    (\Box+(\nabla \ln(G_\Box))\cdot\nabla+\xi R)F_\xi=-\bd_\sigma F_\xi
\eeq
where $\nabla\ln G_\Box\cdot\nabla F_\xi$ is short-hand for
\beq
      \nabla\ln G_\Box\cdot\nabla F_\xi \equiv \frac{1}{G_\Box\sqrt{g}}
      \partial_\mu(\sqrt{g}g^{\mu\nu})(\partial_\nu (G_\Box F_\xi)
      +2g^{\mu\nu}\partial_\mu \ln G_\Box\partial\nu F_\xi \label{eq:h1}
\eeq
By considering the $\xi R$-terms as a kind of mass term, one would expect
$F_\xi$ to be of the form $\exp(-\xi R\sigma+...)$. We will write
\beq
    F_\xi = e^{T_\xi} \label{eq:ft}
\eeq
and then Taylor expand $T_\xi$ in powers of $\sigma$, i.e.
\beq
    T_\xi = \sum_{n=0}^\infty \tau_n \sigma^n \label{eq:expansion1}
\eeq
The boundary condition implies $\tau_0\equiv 0$, and we can arrive at a
recursion relation for the remaining coefficients by inserting 
(\ref{eq:ft}) and (\ref{eq:expansion1}) into (\ref{eq:h1}). But before we can 
collect powers of $\sigma$ we
must also Taylor expand $\nabla\ln G_\Box$ as this also depends on $\sigma$.
Thus, we must write
\beq
    \nabla^a\ln G_\Box = \sum_{n=0}^\infty {\cal G}_n^a\sigma^n
    \label{eq:heatxp}
\eeq
We then obtain
\beqa
    -\sum n\sigma^{n-1}\tau_n &=& 2\sum_{n,m}({\cal G}^a_n\nabla_a\tau_m)
    \sigma^{n+m}+\sum_{n,m}(\nabla^a\tau_n)(\nabla_a\tau_m)
    \sigma^{n+m}+\nonumber\\
    &&\hspace{55mm}\sum_n(\Box\tau_n)\sigma^n+\xi R
\eeqa
i.e.
\beq
    \tau_{n+1}=-\frac{1}{n+1}\left[2\sum_{n'=0}^n\left({\cal G}_{n'}^a\nabla_a
      \tau_{n-n'}
    +(\nabla^a\tau_{n'})(\nabla_a\tau_{n-n'})\right)+\Box_0\tau_n\right] 
      \qquad n\geq 1
\eeq    
which leads to the following first few coefficients:
\beqas
    \tau_0 &\equiv & 0\\
    \tau_1 &=& -\xi R\\
    \tau_2 &=& \xi{\cal G}_0^a\nabla_a R+\frac{1}{2}\xi\Box R\\
    \tau_3 &=& -\frac{2}{3}\xi{\cal G}_0^a\nabla_a\left({\cal G}_0^b \nabla_b
    R+\frac{1}{2}\Box R\right)+\nonumber\\
    &&\frac{1}{3}\xi{\cal G}_1^a\nabla_a R-\frac{1}{3}\xi^2(\nabla R)^2-
    \nonumber\\
    &&\frac{1}{3}\xi\Box\left({\cal G}_0^a\nabla_a R+\frac{1}{2}\Box R\right)
\eeqas
and so on. We notice first of all that $\tau_n$ contains higher and higher
derivatives of $R$ as $n$ grows. We furthermore notice that, in accordance with
our expectations, $T_\xi = -\xi R\sigma+O(\sigma^2)$. Many years ago it
was argued by Parker and coworkers that such a term should be present in
a resummed heat-kernel, see e.g. \cite{JP}, but probably due to technical
difficulties this suggestion doesn't seem to have been taken up by other
authors. By exponentiating not just the scalar curvature but also the 
remaining corrections to the heat kernel of the d'Alembertian in the way we
do it here, we effectively circumvent those technical difficulties as will
be apparent from what follows below.\\
To be able to actually evaluate these coefficients, we need to know ${\cal
G}_n^a$. One easily sees (by inserting the spectral decomposition of the heat 
kernel into equation (\ref{eq:heatxp}) and putting $\sigma=0$) that
\beq
    {\cal G}_0^a = \nabla^a\ln\sum_\lambda \psi_\lambda(x)\psi_\lambda^*(x')
\eeq
where $\psi_\lambda$ denotes the eigenfunctions of $\Box$. Now, this sum is
actually a sum over projections when $x=x'$. As the eigenfunctions are
supposedly complete this means that the sum equals unity, whereby
\beq
    {\cal G}_0^a = 0
\eeq
The next coefficient is found by differentiating once with respect to $\sigma$ 
and then putting $\sigma$ to zero. It is
\beq
    {\cal G}_1^a = \nabla^a\left(\frac{\sum_\lambda \lambda \psi_\lambda(x)
    \psi_\lambda^*(x')}{\sum_\lambda\psi_\lambda(x)\psi_\lambda^*(x')}\right)
\eeq
Again, along the diagonal $x=x'$ the denominator is unity. The numerator
is actually a spectral decomposition of the
d'Alembertian, and hence, inside an integral (such as that appearing in the
definition of the $\zeta$-function), we can substitute $\nabla^a\Box$
for ${\cal G}_1^a$ (possibly up to boundary terms, which we will simply throw
away).\\ 
>From a physical point of view, it is reasonable to only include first 
and second
derivatives of the scalar curvature $R$; higher derivatives would be higher
order quantum effects of the gravitational background, and we are only 
treating a semi-classical approximation
to quantum gravity. With higher order derivatives of $R$ discarded we thus have
\beqa
    \tau_3 &\approx & -\frac{1}{3}\xi^2(\nabla R)^2\\
    \tau_n &\approx & 0\qquad n\geq 4
\eeqa
leading finally to the formula
\beq
      G_{\Box+\xi R}(x,x,\sigma) \approx G_\Box(x,x,\sigma)e^{-\xi R\sigma
      +\frac{1}{2}\xi\Box R\sigma^2-
      \frac{1}{3}\xi^2(\nabla R)^2\sigma^3} \label{eq:nonmin}
\eeq
A very useful formula indeed, allowing us to express the heat kernel
of the non-minimally coupled scalar field in terms of that of the much
simpler minimally coupled field. 
Also, in physical terms, one can think of the first term in the exponent
as due to the background field and the next two terms as due to higher order
corrections of the gravitational field. Thus, as general relativity is only
applicable to the 1 loop level, it is probably beyond the theoretical framework
employed in this paper to expand the exponent to even higher order.\\
The equation (\ref{eq:nonmin}) is actually much more general than we need
for the Friedman-Robertson-Walker geometry, as in this case the 
curvature depends only upon time. In any case, the trick of writing $F_\xi$
as an exponential and the way we were able to eliminate certain derivatives
was what allowed us to be able to find this result generalising the old
result by Jack and Parker \cite{JP}.

\subsection{Effective Action and the Einstein Equations in 
Friedman-Robertson-Walker Space-Time for the Non-Minimally Coupled 
Scalar Field}
In order to find the free energy of a non-minimally coupled scalar field, we
must first obtain the $\zeta$-function, i.e. we must perform an
integration over the fictitious parameter $\sigma$. Using the formula derived
above, the task is to evaluate the following integral
\beqa
    \zeta_{\Box+\xi R}(s) &=& \frac{1}{\Gamma(s)}\int_0^\infty \sigma^{s-1}
    \int G_\Box(x,x,\sigma)e^{-\xi R\sigma+\frac{1}{2}\xi\Box R\sigma^2
    -\frac{1}{3}\xi^2(\nabla R)^2\sigma^3}d^4x d\sigma\nonumber\\
    &=& \frac{1}{\Gamma(s)}\int_0^\infty \sigma^{s-1}\sum_\lambda \int
    |\psi_\lambda({\bf x})|^2h_\lambda^{-1/2}(t)e^{-2\int_0^t\sqrt{h_\lambda(t')
    }dt'}\nonumber\\
    &&\vspace{-10mm}\hspace{60mm}\times e^{-(\xi R+\lambda)\sigma+\frac{1}{2}\xi\Box 
      R\sigma^2-\frac{1}{3}
    \xi^2(\nabla R)^2\sigma^3}d^4xd\sigma\nonumber\\
    &=& \frac{1}{\Gamma(s)}\int_0^\infty\sigma^{s-1}\sum_\lambda\int
    h_\lambda^{-1/2}(t)e^{-2\int_0^t\sqrt{h_\lambda(t')}dt'-(\xi R+\lambda)
    \sigma+\frac{1}{2}\xi\ddot{R}\sigma^2-\frac{1}{3}\xi^2\dot{R}^2\sigma^3}
    dtd\sigma\nonumber\\
	\label{eq:zetanm}
\eeqa
where we have used the fact that $R$ only depends on the time $t$ and not 
on any
of the other coordinates together with the orthonormality of $\psi_\lambda({\bf
x})$ to carry out the integrals over $\chi$ and the angles.
Unfortunately we have not been able to perform this $\sigma$-integration
explicitly, so instead we have to make do with an approximation. The expression
appearing in the exponential has an easy interpretation: the first term is the
flat space-time contribution, the second is the non-minimal coupling term 
while the
remaining two terms are first loop quantum corrections to the gravitational 
background. It thus makes sense to only 
expand these latter terms to the first order, whereby equation 
(\ref{eq:zetanm}) becomes
\beqa
    \zeta_{\Box+\xi R}(s)&\approx&\frac{1}{\Gamma(s)}\sum_\lambda\int 
	dt\int_0^\infty
       d\sigma\sigma^{s-1}e^{-(\xi R+\lambda) \sigma}
    \left(1+\frac{1}{2}\xi\ddot{R}\sigma^2 -\frac{1}{3}
    \xi^2\dot{R}^2\sigma^3\right)h_\lambda^{-1/2}e^{-2\int_0^t\sqrt{h_\lambda
      (t')}dt'}\nonumber\\
    &=&\sum_\lambda\int h_\lambda^{-1/2}e^{-2\int_0^t
    \sqrt{h_\lambda}dt'}\left((\xi R+\lambda)^{-s}+
    \frac{1}{2}\xi\ddot{R}s(s+1)(\xi R+\lambda)^{-s-2}\right.\nonumber\\
    &&\vspace{-1mm}\hspace{50mm}\left.
    -\frac{1}{3}\xi^2\dot{R}^2s(s+1)(s+2)(\xi R+\lambda)^{-s-3}\right)dt
\eeqa
Whence the effective action becomes
\beqa
\hspace{-20mm}\Gamma_{\rm non-minimal} &=& 
    \zeta'(0)\\
    &=&-\sum_\lambda\int h_\lambda^{-1/2}
    e^{-2\int_0^t\sqrt{h_\lambda}dt'}\left( -\ln(\xi R+\lambda) +\frac{1}{2}\xi
    \ddot{R}(\xi R+\lambda)^{-2}\right.\nonumber\\
    &&\hspace{72mm}\left.-\frac{2}{3}\xi^2\dot{R}^2 (\xi R+\lambda)^{-3}
	\right)dt
\eeqa
where the curvature scalar and its derivatives are found to be
\beqa
    R &=& 6a^{-2}(a\ddot{a}+\dot{a}^2+K)\\
    \dot{R} &=& -2a^{-1}\dot{a}R+6a^{-2}(3\dot{a}\ddot{a}+aa^{(3)})\\
    \ddot{R} &=& (6a^{-2}\dot{a}^2-2a^{-1}\ddot{a})R-24
    a^{-3}\dot{a}(3\dot{a}\ddot{a}+a a^{(3)})+\nonumber\\
    &&\hspace{30mm}\qquad\qquad 6a^{-2}(3\ddot{a}^2+4\dot{a}a^{(3)}+aa^{(4)})
\eeqa
here $a^{(3)}$ and $ a^{(4)}$ denotes the third and fourth derivative of
$a$ with respect to time.
With this the pressure, which once more is to enter the Einstein equations as
$\langle T_{33}\rangle$, becomes
\beqa
    p &=& \frac{1}{a^2}\sum_\lambda\left[\left(-\frac{1}{2}h_\lambda^{-3/2}
    \frac{\partial h_\lambda}{\partial a}-\int_0^th_\lambda^{-1/2}
    \frac{\partial h_\lambda}{\partial a}dt'\right)\left(-\ln(\xi R+\lambda)
    \right.\right.\nonumber\\
    &&\qquad \left.+\frac{1}{2}\xi\ddot{R}(\xi R+\lambda)^{-2}-\frac{2}{3}\xi^2
    \dot{R}^2(\xi R+\lambda)^{-3}\right)\nonumber\\
    &&\qquad +h_\lambda^{-1/2}\left(-(\xi R+\lambda)^{-1}\xi\frac{\partial R}{
      \partial
    a}+\frac{1}{2}\xi\frac{\partial\ddot{R}}{\partial a}(\xi R+\lambda)^{-2}
    \right.\nonumber\\
    &&\qquad -\xi^2\ddot{R}(\xi R+\lambda)^{-3}\frac{\partial R}{\partial a}-
    \frac{4}{3}\xi^2\dot{R}\frac{\partial\dot{R}}{\partial a} (\xi R+\lambda)
    ^{-3}\nonumber\\
    &&\qquad \left.\left.+2\xi^3\dot{R}^2(\xi R+\lambda)^{-4}\frac{\partial R}
    {\partial a}\right)\right]e^{-2\int_0^t\sqrt{h_\lambda}dt'}
\eeqa
The differentiations with respect to the scale factor appearing in this
expression are
\beqa
    \frac{\partial R}{\partial a} &=& -2a^{-1}R+6a^{-2}\ddot{a}\\
    \frac{\partial\dot{R}}{\partial a} &=& 6 a^{-2}\dot{a}
    R-12a^{-3}\dot{a}\ddot{a}\\
    \frac{\partial \ddot{R}}{\partial a} &=& -(12
    a^{-3}\dot{a}^2-2a^{-2}\ddot{a})R+(6a^{-2}\dot{a}^2-2a^{-1}\ddot{a})
    (-2a^{-1}R+6a^{-2}\ddot{a})\nonumber\\
    &&\qquad +72a^{-4}\dot{a}(3\dot{a}\ddot{a}+aa^{(3)})-
    12a^{-3}(3\ddot{a}^2+4\dot{a}a^{(3)}+aa^{(4)})\nonumber\\
    &&\qquad -24a^{-3}\dot{a}a^{(3)}+6a^{-2}a^{(4)}    
\eeqa
and
\beq
    \frac{\partial h_\lambda}{\partial a} = -6\dot{a}a^{-3}+\frac{3}{2}
    a^{-2}\ddot{a}-2a^{-3}\lambda
\eeq
All this is to be inserted into the Einstein equations
\beq
    a^{-2}(2\ddot{a}a+\dot{a}^2+K)=p
\eeq
The reader will no doubt appreciate that these equations are not presented in 
their discretised form.\\
Again, as with the minimal coupled case, this result differs from the equations
of motion found by Suen and Anderson \cite{SA} due to the use of the 
WKB-approximation, which speeds up convergence. But this time, there is also
a discrepancy due to the presence of an $e^{-\xi R\sigma}$ term in the
heat kernel, which, as demonstrated elsewhere, should be present \cite{JP}.

\subsection{Performing the Simulations}
We have carried out the same discretisation procedure as for the minimally
coupled case and have simulated the evolution of the various 
Friedman-Robertson-Walker
geometries for the case of conformal coupling $\xi=\frac{1}{6}$. The results 
are shown in figures 2a-c.
\\
We see that this conformal coupling to the gravitational background 
makes the tendencies already
inherent in the minimally coupled case much more significant. For $K=+1$, the
closed universe, the tendency towards collapse is now even more pronounced,
in fact all the chosen values gave rise to a collapse, whereas the 
expansion in the open and flat universes is seen to be faster.\\
Again, for large times, we get power-law behaviour (assuming no surprises for
really large times), giving powers $\sim 0.6$ for $K=0$ and $1.5$ for $K=-1$.
Note that this means that 
for a vacuous FRW-geometry with $K=-1$ and with a conformal scalar field 
as its only inhabitant, the age of the universe is no longer bounded by its 
Hubble time. Also note that for $K=0$ our result deviates only little from that
of Starobinsky (who finds a power of $2/3$) \cite{Star}. 
\\
Let us conclude this section by comparing the evolution for the minimally  
and the conformally coupled scalar field cases, see figure 3, to gain
a little intuition about the influence of the coupling to the background.
\\
It is seen that the coupling to the background tends to speed up 
expansion/collapse
and also that the onset of expansion/collapse occurs earlier. Also some some 
initial conditions that for $K=0$ (seemingly) give rise to expansion in the 
minimally
coupled case gives rise to collapse in the conformal case and {\em vice versa}
some 
initial conditions that for $K=-1$ give rise to collapse in the minimally
coupled case gives rise to expansion in the conformal case.

\section{Relating the Casimir Effect of (Spin 1) Abelian Gauge Bosons 
               to That of a Scalar Field}
In order to be able to actually carry out the computations for higher spins
as well (and thereby eventually being able to handle realistic field
theories such as the standard models or various GUTs), 
we want to derive some relationships between the heat kernels, and
thus the $\zeta$-functions, for vector bosons and Dirac fermions on the 
one hand to
that of the scalar field case on the other. As the case of gauge bosons 
carries some
resemblance to that of the non-minimally coupled scalar field we first
consider that case.\\
We can derive an
expression for the effective action for a spin-1 field, even in the 
non-Abelian case, by a simple extension of the technique used for the 
non-minimally
coupled scalar field, but we need a way to handle the self-interaction (the
non-linear, non-Abelian terms in the field strength tensor), and we will
introduce a way of determining meanfields that
allows us to handle such non-linearities (conseqeuntly,
we could then also handle, say, $\phi^4$ theory in this manner).
\\
The self-interaction is treated by a mean field
approximation -- the mean field itself being expressible in terms of heat
kernels. Due to the implicit complexity of the solution we have not been able 
to carry out the simulations for vector fields, however. But it is important
to emphasise that this trick of using a mean field coupled to the way we
compute heat kernels for non-minimally coupled scalar fields, actually allows
us to find an expression for the heat kernel of a Yang-Mills theory, and
consequently also the effective action of non-Abelian theories.
\\
In order to keep the notation as simple as possible we have left out the ghost
contribution coming from the over-counting of degrees of freedom in the 
functional
integral. In Lorentz gauge, the ghost contribution is simply $-2$ times the
effective action for a minimally coupled scalar-field. It is thus 
straightforward to include this correction at the end.\\
When considering a Yang-Mills field in a curved space-time then, in order
to obtain the field strength tensor the naive guess is to replace the
derivatives of the Minkowski space field strength tensor with covariant derivatives
which is not correct as this leads to a non-gauge covariant expression
(even though, accidentally, it gives the right answer in the case of abelian
fields and no back-ground torsion). Instead proceed by considering the full theory of Dirac fermions
interacting minimally with the gauge fields as well as with the gravitational
field. In order to preserve local gauge and Lorentz covariance construct 
a covariant derivative of the form \cite{prd,Ramond}
\beq
D_m=e^\mu_m(\bd_\mu+\frac{i}{2}\omega^{pq}_\mu(x)X_{pq}+igA^a_\mu(x)T_a)
    \label{eq:covdiv}
\eeq
where $e^\mu_m $ is the vierbein (local base vectors of a freely falling observer), 
$\omega^{pq}_\mu(x)$
is the spin connection being the gravitational analogue of the gauge field
$A^a_\mu(x)$ and $X_{pq}$ the
corresponding Lorentz group ($SO(3,1)$) generators analogous to the
gauge group generators $T_a$. Greek indices refer to curvilinear coordinates
while Latin indices refer to local Lorentz coordinates (of a freely falling observer)
and are also used
for gauge group indices (it should be clear from context: in general we will use
small Latin letters from the beginning of the alphabet to denote gauge indices,
whilst reserving letters from the last half of the alphabet for use as Lorentz 
indices).\\
As in flat space field theory the gauge field strength tensor $F_{mn}^a$ is 
obtained from the commutator of the covariant derivatives
\beq
[D_m,D_n]=S_{mn}^q(x)D_q+\frac{i}{2}R_{mn}^{~~~pq}(x)X_{pq}+iF_{mn}^aT_a
\eeq
(where $S_{mn}^q(x)$ is the torsion and $R_{mn}^{~~~pq}(x)$ is the Riemann
curvature tensor). Using the commutation relations
\beqa
\left[ \bd_\mu,\bd_\nu \right]&=&0 \qquad\mbox{      ,alternatively      }
\qquad\left[ \bd_m,\bd_n \right]=(\bd_me^\mu_n-\bd_ne^\mu_m)e^p_\mu\bd_p
    \label{eq:comrel}\\
\left[ T_a,T_b \right] &=& if_{abc}T^c\\
\left[ X_{mn},X_{pq} \right] 
&=& -i\eta_{mp}X_{nq}+i\eta_{np}X_{mq}-i\eta_{nq}X_{mp}-i\eta_{mq}X_{np}
\eeqa
($\eta_{mn}$ being the flat metric) the normalisations
\beqa
Tr(T_aT_b)&=&\frac{1}{2}\delta_{ab}\\
Tr(X_{mn}X_{pq})&=&-16(\eta_{mp}\eta_{nq}-\eta_{mq}\eta_{np})\label{eq:norm}
\eeqa
and the action of the $SO(3,1)$ generators upon the latin index $m$
\beq
X_{pq}e^\rho_m=i\eta_{qm}e^\rho_q-i\eta_{pm}e^\rho_q
\eeq
one obtains (after a lengthy calculation)
the field strength tensor
\beq
    F_{mn}^a = e_m^\mu e_n^\nu(\partial_\mu A_\nu^a-\partial_\nu A_\mu^a
    +igf_{bc}^a A_\mu^b A_\nu^c) 
\eeq
\vspace{10pt}\par
The generating functional of the full theory of a fermion interacting 
with a non-abelian gauge field is
\beqa
Z&=&\int {\cal D}A_\mu \int D\psi D\bar{\psi}e^{S_{\rm gauge field}+S_{\rm fermion}}\\
&=&\int {\cal D}A_\mu \int D\psi D\bar{\psi}e^{-\frac{1}{4}\int F_{mn}^a F_a^{mn}dx^\mu
                                       +i\int \bar{\psi}\gamma^m D_m\psi dx^\mu}
\eeqa
Referring back to equation (\ref{eq:covdiv}) we see that the fermion part 
of the action
contain reference to the gauge field so that one {\em a priori} cannot
carry out the two integrations independently. One could probably treat 
the fermion action as a source term when doing the gauge field integration
and then subsequently do the fermion integration. Instead however, we are
going to make a mean field approximation to $A_\mu$ in the fermionic 
integral, as well as in the higher order terms of the bosonic integral
(see later). Thus we can consider the bosonic and the fermionic parts
independently (the meanfield in the fermionic path integral is just a function
of space-time variables).
\\
The bosonic part of the generating functional then becomes:
\beqa
Z&=&\int {\cal D}A_\mu e^{-\frac{1}{4}\int F_{mn}^a F_a^{mn}dx^\mu}\nonumber\\
&=&\int {\cal D}A_\mu e^{-\frac{g^2}{4}\int d^4x
       e_m^\alpha e_n^\beta(\partial_\alpha A_\beta^b T_b
                        -\partial_\beta A_\alpha^a T_a
                    -gf_{abc} A_\alpha^a A_\beta^b T^c)
       e_p^\delta e_q^\eta(\partial_\delta A_\eta^e T_e
                        -\partial_\eta A_\delta^d T_d
                    -gf_{def} A_\delta^d A_\eta^e T^f)}\nonumber\\
&=&\int {\cal D}A_\mu \exp(-\frac{g^2}{4}\int d^4x
       \left[\bd_m A_n^bT_b\bd_pA^e_qT_e-\bd_mA^b_nT_b\bd_qA^d_pT_d
              -gf_{def}\bd_mA_n^bT_bA^d_pA^e_qT^f\right.\nonumber\\
      &&\hspace{40mm}-\bd_nA_m^aT_a\bd_pA_q^eT_e+\bd_nA^a_mT_a\bd_qA^d_pT_d
              +gf_{def}\bd_nA^a_mT_aA^d_pA^e_qT^f\nonumber\\
      &&\left.\hspace{10mm}  -gf_{abc}A^a_mA^b_nT^c\bd_pA_q^eT_e
              +gf_{abc}A^a_mA^b_nT^c\bd_qA^d_pT_d
              +g^2f_{abc}f_{def}A^a_mA^b_nT^cA^d_pA^e_qT^f\right]
                               \eta^{pm}\eta^{qn})\nonumber\\
&=&\int {\cal D}A_\mu \exp(-\frac{g^2}{2}\int d^4x
                   \left[A^a_nT_a\bd_m(\bd^nA^{mb}-\bd^mA^{nb})T_b
                        \right.\nonumber\\
    &&\hspace{60mm}+gf_{def}A^d_mA^e_nT^f(\bd^nA^{ma}-\bd^mA^{na})T_a\nonumber\\
            &&\hspace{60mm} \left.   +\frac{1}{2}g^2f_{abc}f_{def}A^a_mA^b_nT^cA^{md}A^{ne}T^f\right])
\eeqa
Using the commutation relation (\ref{eq:comrel}) and the normalisation 
(\ref{eq:norm}) this can be rewritten as 
\beqa
 &&\int {\cal D}A_\mu \exp(-\frac{g^2}{4}\int d^4x
   \left[A^a_n(( \bd_me^\mu_n-\bd_ne^\mu_m ) e^p_\mu\bd_p+\bd^n\bd_m)A^m_a
          -A^a_n\bd_m\bd^mA^n_a\right.\nonumber\\
 &&\left.    +gf_{dea}A^d_mA^e_n(\bd^nA^{ma}-\bd^mA^{na})
          +\frac{1}{2}g^2f_{abc}f_{def}A^a_mA^{md}A^b_nA^{ne}\right])
\eeqa
where the last part of the first term can be eliminated by applying the
Lorentz condition, $\bd_mA^m_a=0$. Now make the following mean field
approximation to the path integral:
\beqa
Z&=& \int {\cal D}A_\mu \exp(-\frac{g^2}{4}\int d^4x
   \left[A^a_n(\bd_me^\mu_n-\bd_ne^\mu_m)e^p_\mu\bd_pA^m_a
          -A^a_n\bd_m\bd^mA^n_a\right.\nonumber\\
  &&\hspace{35mm}\left. +A^e_n\langle gf_{dea}(\bd^nA^{ma}-\bd^mA^{na})\rangle A^d_m+A^{ne}
    \langle\frac{1}{2}g^2f_{abc}f_{def}A^a_mA^b_n\rangle A^{md}\right])\nonumber\\
&=&\hspace{-5mm}\int {\cal D}A_\mu e^{-\int d^4x
 A^b_m\frac{g^2}{4}\left[-\delta^a_b\delta^m_n\bd_p\bd^p
         +\delta^a_b(\bd_ne^{m\mu}-\bd^me^\mu_n)e^p_\mu\bd_p
         +<gf_{b\hspace{3pt}c}^{\hspace{3pt}a}(\bd_nA^{mc}-\bd^mA^c_n)>
+<\frac{1}{2}\delta^m_ng^2f_{eb}^{~~c}f_{d\hspace{3pt}c}^{\hspace{3pt}a}A^e_pA^{pd}>
               \right]A^n_a}\nonumber\\
\eeqa
\subsection{Solving the Heat Kernel Equation}
The full heat kernel equation then is
\beqa
&&\hspace{-10mm}\frac{g^2}{4}\left[\delta^a_b\delta^m_r\bd_p\bd^p
         -\delta^a_b(\bd_re^{m\mu}-\bd^me^\mu_r)e^p_\mu\bd_p
         -\langle gf_{b\hspace{3pt}c}^{\hspace{3pt}a}(\bd_rA^{mc}-\bd^mA^c_r)\rangle\right.\nonumber\\
&&\left.-\langle\frac{1}{2}\delta^m_rg^2f_{ebc}f_{d\hspace{3pt}c}^{\hspace{3pt}a}A^e_pA^{pd}\rangle
               \right] G_{n(b)}^{r(a)}(x,x',\sigma)=-\bd_\sigma G_{n(b)}^{m(a)}(x,x',\sigma) \label{eq:heat1}
\eeqa
or, in short hand notation
\beq
\frac{g^2}{4}\left[\delta^a_b\delta^m_k\Box_0
         -\delta^a_b{\cal E}^{mp}_k\bd_p
         -f_{k(b)}^{m(a)}(\langle A\rangle)      \right] G_{n(b)}^{k(a)}(x,x',\sigma)=-\bd_\sigma G_{n(b)}^{m(a)}(x,x',\sigma)
    \label{eq:spin1}
\eeq
with\footnote{
The quantity ${\cal E}_{mn}^p$ is the structure coefficients of the algebra
of ``flat'' derivatives $\partial_m$. Thus it is therefore not surprising that
it occurs here.}
\beq
    {\cal E}_n^{mp} = \left(\partial_ne^{m\mu}-\partial^me_n^\mu\right) e^p_\mu
\eeq
Now, in order to be able to make use of the WKB solution from the scalar case, we must replace the
``flat'' d'Alembertian $\Box_0 \equiv \bd_p\bd^p$, by the proper d'Alembertian operator on a scalar field, $\Box$. From the
definition (\ref{eq:def-box}) with $g_{\mu\nu}=\eta_{ab}e^a_\mu e^b_\nu$ and $e=\sqrt{g}$ being the vierbein determinant, 
it follows that
\beq
    \Box_0 = \Box - \frac{1}{e}\bd_\mu(ee^\mu_p)\bd^p   \label{eq:box0}
\eeq
Inserting this into (\ref{eq:spin1}) we get
\beq
    \left[\delta^a_b\delta^m_k\Box - \delta^a_b\bar{\cal E}^{mp}_k\bd_p -f_{k(b)}^{m(a)}(\langle A\rangle)\right]
    G_{n(b)}^{k(a)}(x,x',\bar{\sigma}) = -\bd_\sigma G_{n(b)}^{m(a)}(x,x',\bar{\sigma}) \label{eq:heat1}
\eeq
where we have also rescaled $\sigma\rightarrow \bar{\sigma}=
g^{2}\sigma/4$, in order to get rid of the over all factor of
$\frac{g^2}{4}$ in ({\ref{eq:spin1}) and also introduced $\bar{\cal E}$ 
defined by
\beq
 \bar{\cal E}^{mp}_k =  {\cal E}^{mp}_k + \frac{1}{e}\bd_\mu (e e^{\mu~p}\delta_k^m)  
\eeq
\par
The first order term is eliminated, as before, by the substitution 
\beq
    G = \tilde{G} e^{\frac{1}{2}\int \bar{\cal E}^{mp}_n dx_p} \label{eq:non-cov}
\eeq
where $G$ is a tensor but the factors on the right side of the 
equation independently seen are not,
because the exponent is not. This leads to the following equation
\beq
\left[\delta^a_b\delta^m_r\Box
         +\frac{1}{2}\bd_p\bar{\cal E}^{mp}_n+\frac{1}{4}\bar{\cal E}^{mp}_k
	\bar{\cal E}^k_{np}
    +f^{m(a)}_{r(b)}(\langle A\rangle )
               \right] G_{n(c)}^{r(b)}(x,x',\sigma)=-\bd_\sigma 
	G_{n(c)}^{m(a)}(x,x',\sigma) \label{eq:heat2}
\eeq
which is of the form
\beq
    \delta^a_c\left[\delta^m_r\Box +{\cal O}_{r(c)}^{m(a)}\right]
                      G_{n(b)}^{r(c)}(x,x',\bar{\sigma})
    =-\bd_{\bar{\sigma}} G_{n(b)}^{m(a)}(x,x',\bar{\sigma}) \label{eq:heat3}
\eeq
where $O=O(\langle A(x)\rangle, x)$ is a matrix-valued function of $x$.
The heat kernel becomes (once more) a matrix-valued function which we can 
assume to be of the form
\beq
    G_{n(b)}^{m(a)}(x,x',\bar{\sigma}) = G_\Box(x,x',\bar{\sigma}) (e^{{\sf T}(x,x'\bar{\sigma})})_{n(b)}^{m(a)}
\eeq
where $G_\Box$ denotes the heat-kernel of $\Box$ and $\sf T$ is some matrix 
$({\sf T})_{n(b)}^{m(a)}=T_{n(b)}^{m(a)}$. Inserting this expression for the heat-kernel into the
heat equation we arrive at an equation for $T_{n(b)}^{m(a)}$
\beq
    \Box T_{n(b)}^{m(a)} + (\partial_p T^{m(a)}_{k(c)})(\partial^pT_{n(b)}^{k(c)})-2\bd_p\ln(G_\Box)\bd^pT^{m(a)}_{n(b)}+
    {\cal O}_{n(b)}^{m(a)} = 
    -\frac{\partial}{\partial\bar{\sigma}}T_{n(b)}^{m(a)} \label{eq:Teq}
\eeq
where a summation over repeated indices is understood. The third term on the
left hand side vanishes
along the diagonal $x=x'$ by the argument given in section 4. We will furthermore write $T_{n(b)}^{m(a)}$ as a Taylor
series
\beq
    T_{n(b)}^{m(a)}(x,\bar{\sigma}) = \sum_{\nu=0}^\infty \tau_{~~~n(b)}^{(\nu)m(a)}(x)\bar{\sigma}^\nu \label{eq:expansion}
\eeq
This leads to a recursion relation for the coefficients $\tau_{~~~n(b)}^{(\nu)m(a)}$
\beq
    \Box\tau_{~~~n(b)}^{(\nu)m(a)}+\sum_{\nu'=0}^\nu(\partial_p\tau_{~~~~~~k(c)}^{(\nu-\nu')m(a)})
    (\partial^p\tau_{~~~n(b)}^{(\nu')k(c)}) = -(\nu+1)\tau_{~~~n(b)}^{(\nu)m(a)}  \label{eq:recursion}
\eeq
or, as $\bd_p=e^\mu_p \bd_\mu$ and $e^\mu_p$ is the inverse of $e^p_\mu$:
\beq
    \Box\tau_{~~~n(b)}^{(\nu)m(a)}+\sum_{\nu'=0}^\nu(\partial_p\tau_{~~~~~~k(c)}^{(\nu-\nu')m(a)})
    (\partial^p\tau_{~~~n(b)}^{(\nu')k(c)}) = -(\nu+1)\tau_{~~~n(b)}^{(\nu)m(a)}
\eeq
\vspace{15pt}
\par
Due to the boundary condition one has 
\beq
\bd^p\tau_{~~~~n(b)}^{(0)~m(a)}=\frac{1}{2}\delta^a_b \bar{\cal E}^{mp}_n
\eeq
and because of this and equation (\ref{eq:Teq}) the next coefficent becomes
\beq
\tau^{(1)~m(a)}_{~~n(b)}= \delta_a^b\bd_p\bar{\cal E}^{mp}_n-\frac{1}{2}\delta^a_b\eta_{pq}
    \bar{\cal E}_k^{mp}\bar{\cal E}_n^{kq}+f_{n(b)}^{m(a)}(\langle A\rangle)   
\eeq
and subsequently, cf equation (\ref{eq:recursion}), the next few coefficients 
are seen to be  
\beqa
\tau^{(2)~m(a)}_{~~~~n(b)}&=& -\frac{1}{2}\delta^a_b\Box\bd_p\bar{\cal E}_n^{mp}\nonumber\\
&&+\frac{1}{4}\delta^a_b\Box(\bar{\cal E}_k^{mp}\bar{\cal E}_n^{kq}\eta_{pq})\nonumber\\
&&+\frac{1}{4}\delta^a_b(\bd_q\bd_p\bar{\cal E}_k^{mp})\bar{\cal E}_n^{kq}\nonumber\\
&&-\frac{1}{8}\delta^a_b\bd_q(\bar{\cal E}_k^{mp}\bar{\cal E}_l^{kr}\eta_{pr})\bar{\cal E}_n^{lq}\nonumber\\
&&+\frac{1}{4}\delta^a_b\bar{\cal E}_k^{mq}\bd_q\bd_p(\bar{\cal E}_n^{kp})\nonumber\\
&&-\frac{1}{8}\delta^a_b\bar{\cal E}_k^{mq} 
                    \bd_q(\bar{\cal E}_k^{lp}\bar{\cal E}_n^{lr}\eta_{pr})\nonumber\\
&&-\frac{1}{2}\Box f_{n(b)}^{m(a)}(\langle A\rangle)                \nonumber\\
&&-\frac{1}{2}\bd_p f_{k(b)}^{m(a)}(\langle A\rangle) \bar{\cal E}_n^{kp}\nonumber\\
&&-\frac{1}{2}\bar{\cal E}_k^{mp}\bd_p f_{n(b)}^{k(a)}(\langle A\rangle) \nonumber\\
&\approx&-\frac{1}{2}\Box f_{n(b)}^{m(a)}(\langle A\rangle)                \nonumber\\
&&-\frac{1}{2}\bd_p f_{k(b)}^{m(a)}(\langle A\rangle) \bar{\cal E}_n^{kp}\nonumber\\
&&-\frac{1}{2}\bar{\cal E}_k^{mp}\bd_p f_{n(b)}^{k(a)}(\langle A\rangle) 
\eeqa
and
\beqa
\tau^{(3)~m(a)}_{~~~~n(b)}&=&-\frac{1}{3}
\left\{\Box[-\frac{1}{2}\bd_p f^{m(a)}_{k(b)}\bar{\cal E}_n^{kp}\right.\nonumber\\
        && -\frac{1}{2}\bar{\cal E}_n^{kp}\bd_p f_{k(b)}^{m(a)}  ]\nonumber\\
&&-\frac{1}{2}\bd_p\left[-\frac{1}{2}\bd_q f_{k(b)}^{m(a)}(\langle A\rangle) \cdot 
                       \bar{\cal E}_l^{kp}\right]\bar{\cal E}_n^{lq}\nonumber\\
&&-\frac{1}{2}\bar{\cal E}_k^{mp}\bd_p\left[-\frac{1}{2}\bd_q f_{l(b)}^{k(a)}(\langle A\rangle)
    \bar{\cal E}_n^{lp}\right]\nonumber\\
&&+\bd_pf_{k(b)}^{m(c)}\bd^pf^{k(a)}_{n(c)}\nonumber\\
&&-\frac{1}{4}\bd_p\Box f^{m(a)}_{k(b)} \bar{\cal E}_n^{kp}\nonumber\\
&&-\frac{1}{4}\bar{\cal E}_k^{mp}\bd_p\Box f^{k(a)}_{n(b)}\nonumber\\
&&+\bd^qf^{m(a)}_{k(b)}\bd_q \bd_p\bar{\cal E}_n^{kp}\nonumber\\
&&+\bd_q \bd_p\bar{\cal E}_k^{mp}\bd^qf^{k(a)}_{n(b)}\nonumber\\
&&-\frac{1}{2}\left(\bd_q \bar{\cal E}_k^{mp}\right)^2 \bd^qf^{k(a)}_{n(b)}\nonumber\\
  &&\left.-\frac{1}{2}\bd^qf^{m(a)}_{k(b)} 
       \left( \bd_q \bar{\cal E}_n^{kp}\right)^2 \right\}\nonumber\\
    && \mbox{   +  terms of even higher order  }
\nonumber\\
&\approx&\bd_pf^{m(c)}_{k(b)}\bd^pf^{k(a)}_{n(c)}
\eeqa
and so forth.\\
For computational purposes one should, as indicated above, keep only the 
terms that seem most important when doing
simulations, i.e., the most important pure curvature term, the most 
important pure gauge field terms
and the most important mixing/gauge field-curvature coupling terms\footnote{
Note, however, that by discarding terms that are essentially higher and 
higher order derivatives of
the curvature one might lose a large contribution when close to singularities.
Also note the discussion at the end of this subsection.}. 
Thus we arrive at the following expression for the heat kernel: 
\beq
    G_{n(b)}^{m(a)}(x,x,\bar{\sigma}) = G_\Box(x,x,\bar{\sigma})
	\left(e^{\tau^{(0)}\bar{\sigma}-\frac{1}{2}\tau^{(1)}\bar{\sigma}^2
    -\frac{1}{3}\tau^{(3)}\bar{\sigma}^3}\right)_{n(b)}^{m(a)}  
	\label{eq:spin1solution}
\eeq
This formula then expresses the heat kernel for a vector boson in a curved
space-time in terms of that of a scalar field in flat space-time and a
matrix-valued function depending on the curvature.\\
One could go on to develop
a similar formula for the heat kernel for a spin $s$ boson then, instead of a
matrix-valued function, we would then get a rank $2s$ tensor-valued function.
%\\
%When using the above result note should be given to the
%fact that the two terms of equation (\ref{eq:non-cov}) only give a covariant
%result when multiplied. Thus, because of the truncation of the series (\ref{eq:expansion})
%i.e. because one of the terms is only approximation, the above solution will
%in general not be covariant (the severity of this problem being dependent on
%the application at hand). However, one might still expect to get a reasonable estimate
%of the effect of the resident spin 1 fields\footnote{One might even be as 
%optimistic as to note that in many problems involving differential equations 
%the basis for the solutions
%break the symmetry of the problem. For instance spherical harmonics do not 
%posses spherical symmetry.}
%
%
%
\subsection{Determining the Mean Field}
In the above we substituted mean fields for the Yang-Mills field in the
non-abelian terms of the action.
We now proceed to determine these mean fields.
By definition the gauge mean field squared is\footnote{Wherever 
the mean value of a any odd power of the
gauge fields occur, we substitute $\sqrt{\langle A^2\rangle}$ for $\langle A
\rangle$. 
Of course, for $\langle A\rangle\neq 0$, this corresponds to
spontaneous symmetry breaking of gauge symmetry (at least in the meanfield
approximation) due to the fact that virtual particles interact with gravity
while propagating (see e.g. \cite{GMM}). 
A couple of
things have to be said about this fact. Firstly, it has been noted that the
effective action need not be gauge invariant beyond the 1-loop level 
\cite{effact}. Secondly, we do indeed prefer to see the symmetry breaking 
as an indication that a more complete theory would have
to have some `mixing' of gauge transformations and coordinate ones, i.e., one
should only be able consider gauge transformations independently of local 
Lorentz ones as a `corner' of the theory. Actually, at the higher level one 
might well
find that masses occur naturally in the theory, as witnessed by the condensate
formation seen in the mean field approximation. 
Thirdly, it is most satisfying that the fact that an observer in curved space do see an (Casimir) energy 
density
is related to the fact that mean fields do attain finite, non-zero values in
curved space vacuum.}
\beq
    \langle A_m^a(x) A_n^b(x')\rangle \equiv \frac{\int A_m^a(x) A_n^b(x')
    e^{iS}{\cal D}A}{\int e^{iS}{\cal D}A}
\eeq
where $S$ denotes the appropriate action. We want to carry out this functional
integral. In order to do this, we note that in $n$-dimensional Euclidean space
the following holds\footnote{One should note that this gives an interpretation
of the corresponding $\zeta$-function at $s=1$, since we get
$\langle x^2\rangle = \frac{1}{2}\zeta_M(1)$.} (see appendix for derivation)
\beqs
    \langle x_i x_j\rangle \equiv \frac{\int x_ix_je^{-x^tMx}d^nx}{\int
    e^{-x^tMx}d^nx} = \frac{1}{2}\delta_{ij}(M^{-1})_{ii}
\eeqs
where $M$ is some symmetric matrix. 
This result is independent of the
dimensionality $n$ and can hence be carried over to the infinite dimensional
Hilbert space $L^2$. Thus, in the continuum case where we have $M\sim
\frac{\delta^2S}{\delta A_m^a(x)\delta A_n^b(x')}$, we have by inference 
\beq
    \langle A_m^a(x)A_n^b(x')\rangle  = \frac{1}{2}\left(\frac{\delta^2 S}{\delta
    A_m^a(x)\delta A_n^b(x')}\right)^{-1}
    \equiv \frac{1}{2}\delta^{ab}\eta_{mn}\delta(x,x')
    D(x,x')
\eeq
with $D$ being the Green's function -- 
(the inverse of the matrix $M=\frac{\delta^2S}{\delta A^2}$). 
\\
Now, given an operator $A$, a simple
relationship exists between its heat kernel and its corresponding Green's
function, namely
\beq
    D(x,x') = -\int_0^\infty G_A(x,x',\sigma)d\sigma \label{eq:green}
\eeq
We can derive this formula by writing the Green's function (i.e., the left
inverse of the operator, $AD(x,x')=\delta(x,x')$) in terms of the eigenfunctions $\psi_\lambda$ of $A$:
\beq
    D(x,x') = \sum_\lambda \frac{1}{\lambda}\psi_\lambda^*(x')\psi_\lambda(x)
\eeq
and then using
\beqs
    \frac{1}{\lambda} = -\int_0^\infty e^{-\lambda\sigma}d\sigma
\eeqs
and the spectral representation of the heat kernel to obtain (\ref{eq:green}). 
In order to obtain the mean field values $\langle
A_m^aA_n^b\rangle$ we use, to a first approximation, that part of 
the action which only contains
the abelian terms (i.e., excluding the terms involving the structure
constants, $f_{abc}$, of the Lie algebra). If one desires a better 
approximation, the procedure of this paragraph may be repeated using the
full path integral with the first meanfield approximation substituted for 
the gauge fields. This scheme can then be iterated until the desired 
accuracy is obtained. Towards the end of this subsection, we will describe 
the changes being made in the appropriate expressions during this iteration.\\
The heat kernel equation is as equation (\ref{eq:heat2}) but with 
$f_{r(b)}^{m(a)}\equiv 0$.
The solution is then (still in matrix notation) to the chosen order
\beq
    \tilde{G}(x,x',\bar{\sigma}) = G_{\Box_0}(x,x',{\sigma})e^{-{\cal A}
    \bar{\sigma}
    +\frac{1}{2}{\cal B}\bar{\sigma}^2-\frac{1}{3}{\cal C}\bar{\sigma}^3}
\eeq
with (using (\ref{eq:recursion})) the coefficients given by 
(to the order chosen)\footnote{Even though
the matrices $\cal A,B,C$ do not {\em a priori} commute, there is no problem
here to the chosen order of accuracy. It is conceivable, though, 
that one encounters difficulties
when iterating the scheme for determining the meanfield because the
gauge field dependent terms are not symmetric.}
\beqa
    {\cal A}^m_n &=& -\partial_p\bar{\cal E}^{mp}_n+\frac{1}{2}\bar{\cal E}^{mp}_k
    \bar{\cal E}_{np}^k\\
    {\cal B}_n^m &=& \Box{\cal A}_n^m\\
    {\cal C}^m_n &=& (\partial_p {\cal A}^m_k)(\partial^p {\cal A}^k_n)
\eeqa

For the mean field we then get 
\beqa
    \langle A_m^a(x)A_n^b(x')\rangle  
    &=&\lim_{s\rightarrow 0}\frac{1}{2}\delta^{ab}\eta_{mn}\delta(x-x')
    \int_0^\infty d\sigma \sigma^{s-2} G(x,x,\sigma)\nonumber\\
    &=&\lim_{s\rightarrow 0}\frac{1}{2}\delta^{ab}\eta_{mn}\delta(x-x')
    \int_0^\infty d\sigma \sigma^{s-2}e^{-{\cal A}\sigma+\frac{1}{2}{\cal B}\sigma^2
    -\frac{1}{3}{\cal C}\sigma^3}\nonumber\\
    &\approx &(4\pi)^{-2}\frac{1}{2}\delta^{ab}\eta_{mn}\delta(x-x')\lim_{s\rightarrow 0}\int_0^\infty 
    \sigma^{s-2}e^{-{\cal A}\sigma}\left(1+\frac{1}{2}{\cal B}\sigma^2 -
    \frac{1}{3}{\cal C}\sigma^3\right)d\sigma\nonumber\\
    &=& \frac{1}{32\pi^2}\delta^{ab}\eta_{mn}\delta(x-x')\nonumber\\
    &&\hspace{5mm}\times\lim_{s\rightarrow 0}\left({\cal A}^{-(s-1)}\Gamma(s-1) +
    \frac{1}{2}{\cal BA}^{-(s+1)}\Gamma(s+1)-\frac{1}{3}{\cal CA}^{-(s+2)}\Gamma(s+2)\right)\nonumber\\
\eeqa
where we have used that $\cal A$ is related to the
curvature, and hence we can make do with only including the first terms in the
Taylor series defining $\exp(\frac{1}{2}{\cal B}\sigma^2-\frac{1}{3}{\cal C}\sigma^3)$, just
like we did for the non-minimally coupled scalar field in an earlier section. It
turns out, however, that the resulting integral is divergent in our case (as $s\rightarrow 0$ we get a term
involving $\Gamma(-1)$, since we do not have a $1/\Gamma(s)$ factor outside). We regularise this singularity
simply by using the principal value of the meromorphic function $\Gamma(z)$ at
$z=-1$ (see \cite{princip}). Laurent expanding we get
\beqs
    \Gamma(z) = \frac{-1}{z+1}+(\gamma-1)-\frac{1}{12}(z+1)(6\gamma^2 -12\gamma
    +\pi^2+12)+...
\eeqs
Since the principal part of a meromorphic function at a pole is just the finite
part, we finally obtain
\beqa
    \langle A_m^a(x) A_n^b(x)\rangle_{\rm reg} &=& 
    -\delta^{ab}\left(\left[{\cal A}
    (\gamma-1)+\frac{1}{2}{\cal BA}^{-1}-\frac{1}{3}{\cal CA}^{-2}\right]\right)_{mn}
    \\
    &=& -\delta^{ab}\left(({\cal A}_{mp}(\gamma-1) +\frac{1}{2}
    {\cal B}_{mk} ({\cal A}^{-1})^k_p-\frac{1}{3}{\cal C}_{mk}({\cal A}^{-1})^k_q({\cal A}^{-1})^q_p\right)\nonumber\\
\eeqa
Now, upon iteration of this we must obviously include the mean field in 
this. We then take the above result for $\langle A_m^aA_n^b\rangle_{\rm reg}$
and insert it into the action for the Yang-Mills field as an extra term. The
heat kernel for this new action is then found (the mean fields merely 
corresponds to an extra curvature term) by the same methods as above, from
which one gets a better approximation to the mean fields, and so on.

\subsection{Using the method on other problems}
A note on the generality of the above calculation of the mean field is
in order: We removed the flat space d'Alembertian $\Box_0$ by introducing 
the curved space d'Alembertian $\Box$ through equation (\ref{eq:box0}). This
was done to mimic the case of the non-minimally coupled scalar field and
is possible because we know $G_\Box$ from the section of the minimally
coupled scalar field. However, for other problems one might not know $G_{\Box_0}$.
It should thus be remarked that had we continued using $\Box_0$ instead of
$\Box$ the functional form of the heat kernel equation (\ref{eq:heat3}) would
have been unchanged and the derivations proceed as before. One can then 
take advantage of the fact that the heat kernel of 
the 'flat' space-time d'Alembertian, $\Box_0$ , is known to 
be (se e.g.
\cite{prd,Ramond,GMM})\footnote{Actually, the formula in \cite{Ramond} 
is for Cartesian coordinates. Here
we have brought it on a covariant form. This form is the simplest possible 
form for the heat kernel
compatible with the equivalence principle.} 
\beq
    G_{\Box_0}(x,x',\bar{\sigma}) = \frac{1}{(4\pi\bar{\sigma})^2}
                                   e^{-\frac{\Delta(x,x')^2}{4\bar{\sigma}}}
\eeq
in four dimensions. Here $\Delta(x,x')$ is the geodesic distance between the points $x$ and $x'$, i.e. $\Delta=\int_{x'}^x ds$,
which in Cartesian coordinates would be simply $|x-x'|$. The exact form is not needed, only the fact that 
$G_{\Box_0}$ taken along the diagonal $x=x'$ is independent 
of $x$. One should also note that one can only use
freely falling coordinates for as well $x$ as $x'$ when these are 
sufficiently close to each other, i.e., both within some
sufficiently small neighbourhood. But as we would only be interested, 
ultimately, in the limit $x'\rightarrow x$, this approximation would be
justified.

\section{Heat Kernel Equation and Zeta-Function for Spin 1/2 Fermions}
In the fermionic case the heat kernel equation becomes
\beq
\dslash G_{\dslashf}(x,x',\sigma)=
-\frac{\partial }{\partial \sigma }G_{\dslashf}(x,x',\sigma)
\eeq
Now assume that we know the eigenfunctions of $\dslash$ i.e. $\dslash 
(\psi_\lambda)_\alpha =\lambda (\psi_\lambda)_\alpha$
and guess that the heat kernel has the following form
\beq
 G_{\dslashf}(x,x',\sigma)=\sum_\lambda c^{\alpha\beta} 
    (\psi_\lambda(x'))^*_\alpha
    (\psi_\lambda(x))_\beta e^{g(\lambda)\sigma}
\eeq
which inserted into the heat kernel equation gives
\beq
 \dslash G_{\dslashf}(x,x',\sigma)=\sum_\lambda \dslash c^{\alpha\beta} 
(\psi_\lambda(x'))^*_\alpha(\psi_\lambda(x))_\beta e^{g(\lambda)\sigma}
  =-\frac{\partial}{\partial \sigma}G_{\dslashf}(x,x',\sigma)
	\label{eq:heatspin}
\eeq
provided that $[\dslash,c^{\alpha\beta}]=0$ and $g(\lambda)=-\lambda$.\\
To establish the link with the scalar case make exactly the same 
exercise for the operator $\dslash^2$ i.e. assume that the heat kernel for
$\dslash^2$ is
\beq
    G_{\dslashf^2}(x,x',\sigma)=\sum_\lambda c^{\alpha\beta} 
    (\psi_\lambda(x'))^*_\alpha
    (\psi_\lambda(x))_\beta e^{h(\lambda)\sigma}
\eeq
and insert this into the heat kernel equation to get (using $\dslash 
(\psi_\lambda)_\alpha =\lambda (\psi_\lambda)_\alpha$ twice)
\beqa
\dslash^2 G_{\dslashf^2}(x,x',\sigma)
&=&\sum_\lambda \lambda^2 c^{\alpha\beta} (\psi_\lambda(x'))^*_\alpha
(\psi_\lambda(x))_\beta e^{h(\lambda)\sigma} \nonumber\\
&&  =-\frac{\partial}{\partial \sigma}G_{\dslashf^2}(x,x',\sigma) 
    \mbox{  if one choose   }h(x)=-\lambda^2\label{eq:spinnorm}
\eeqa
If one choose to normalise the spinor eigenfunctions as follows\footnote{
choosing another normalisation trivially alters equation (\ref{eq:spinnorm}).}
\beq
\sum_\lambda \int \sqrt{g}d^4x \sqrt{g}d^4x' c^{\alpha\beta} 
(\psi_\lambda(x'))^*_\alpha
(\psi_\lambda'(x))_\beta \delta(x-x')  = \delta(\lambda-\lambda')
\eeq
(if $c^{\alpha\beta}$ is chosen to be the Kronecker delta 
$\delta^{\alpha\beta}$ then this is the ordinary 
normalisation) then
the zeta-functions are related as follows (cf equation (\ref{eq:heatspin}));
\beqa
\zeta_{\dslashf^2} (s)&=& \frac{1}{\Gamma(s)}\int d\sigma \sigma^{s-1}
    \sum_\lambda\int G_{\dslashf^2}(x,x,\sigma)\sqrt{g}d^4x\nonumber\\
           &=&\sum_\lambda\frac{1}{\Gamma(s)}\int d\sigma \sigma^{s-1}
            e^{-\lambda^2\sigma}\nonumber\\
           &=&\sum_\lambda \lambda^{-2s}\nonumber\\
           &=&\zeta_{\dslashf} (2s)
\eeqa

\subsection{Heat Kernel Equation and Zeta-Function for Free Spin 1/2 Fermions}
Now note that for a free fermion field the covariant derivative is
\beq
\dslash=\gamma^m e^\mu_m(\bd_\mu+\frac{i}{2}\omega^{pq}_\mu(x)X_{pq})
\eeq
Representing the $SO(3,1)$ generators in terms of 
the sigma matrices, $\sigma_{pq}=\frac{i}{4}[\gamma_p,\gamma_q]$, 
one obtains for the derivative squared
\beq
\dslash^2=(\Box+\xi_f R)\cdot {\bf 1}_4
\eeq
(where ${\bf 1}_4$ is the four dimensional unit matrix, and $\xi_f=\frac{1}{8}$)
establishing the link between the scalar and the fermion cases if one 
remembers to
include a factor of 4 (one for each spinor component) in the Dirac case
(corresponding to taking the trace over the unit matrix).
Thus we find
\beq
    \zeta_{\dslashf}(s) = \zeta_{\dslashf^2}(\frac{1}{2}s) = 4\zeta_{\Box+\xi_f R}
    ^{\rm scalar}(\frac{s}{2})
\eeq
This is our final result for free Dirac spinors. We have now arrived at formulas
expressing the heat kernels of free higher spin fields in terms of that of the scalar
field. We notice, however, that in the fermion case we arrive at a non-minimally
coupled scalar field equation. The above result is the result we will use when 
performing the simulations. The case of interacting fields is discussed in the 
appendices.

\subsection{Performing the Simulations}
Due to the simple relationship between the $\zeta$-functions for a non-minimally
coupled scalar field and the Dirac field, the simulations of evolution of
Friedman-Robertson-Walker geometries due to vacuum fluctuations of fermionic fields can
be carried out quite easily. As the functional integral giving the free energy
is now over Grassmannian variables, the sign of the energy will be different (we
will get the determinant raised to the power of plus one half and not minus one
half as for bosons), due to the abovementioned simple relationship we will
furthermore get a factor four (one for each spinor component) and a factor $1/2$
from the argument of the $\zeta$-function. These changes, however, are easily
performed and we can carry out the simulations with very little change to the
program for non-minimally coupled scalar fields. 
We ran the simulations with an initial value of the scale factor fixed at $5 L_{Planck}$.
The results are, for $K=0,\pm 1$ plotted in
figure 4. Note that all $K=+1$ universes either collapse or seem to be on 
the verge of
collapse and that the evolution
for $K=0$ and $K=-1$ tends to converge. Again, this latter
part of the expansion is polynomial in time, the corresponding power is $\sim 0.7$.
Also note that rather contrary to what one might expect due to the fact that
the pressure has the opposite sign of that of the case of a (conformal or scalar) scalar
field (cf formulas (\ref{eq:det}) and the corresponding one for Grassmann fields for which the
power is positive), the behaviour is not qualitatively very different in
the two cases. Actually, the fermionic case looks rather similar to the minimally 
coupled scalar field.

\section{Conclusion}
We have derived rather simple expressions for the effective actions for quantum fields of spin zero, 
one-half and one in a Friedman-Robertson-Walker background, thereby 
obtaining the equations of motion governing the gravitational 
degrees of freedom i.e. the scale factor $a(t)$ of a vacuous
FRW universe. Or, in other words, the Einstein equation now
describes the 
coupling of the zero-point energy (or pressure), the so-called
Casimir energy, to the Einstein tensor, i.e. we consider the 
evolution of space-time including back-reaction of gravity
upon itself as mediated through virtual particles of (other)
quantum fields. This is not the whole picture (direct gravitational self-interaction is 
of course not included) but probably is as far seeing gravitation as a background
field (first quantisation) can bring us.\\
Looking at the plots from the simulations we see first of all that, irrespective of whether the
residing field is minimally or conformally coupled scalar field or a Dirac fermion field, the 
closed universe $K=+1$ seems to be ruled out, as the simulations
show a marked tendency to collapse in this case.\\
For $K=0$ there is a tendency towards expansion according to a power law with power $\sim 0.6-0.7$,
whatever the residing field.\\
For $K=-1$ however, the nature of the residing field is rather important: In all cases
there is a marked tendency, at large times, towards expansion according to some power law,
the power being $\sim 0.7$ for fermions, $\sim 0.8$ for a minimally coupled scalar field
and $\sim 1.5$ for a conformally coupled scalar field. \\
This indicates two things: First the evolution of even a vacuous space-time may depend
crucially on its (virtual) content, i.e. on the residing fields. Secondly, the 
quantum field content might conspire to give expansion according to power laws with
powers greater than $1$. Thus, the age of a FRW-universe need not be less than its
Hubble-time.  Or stated differently; the local Hubble expansion could be larger than
the one observed in the early universe (at large redshifts).
\\
It should be noted that this gravitational back-reaction through quantum fields 
cannot be done away with, it should be included even in the simplest of models that
attempts to describe the real universe. We failed to make simulations
for the case where all fields of the standard model resides in the FRW-geometry
so we abstain from guessing at the evolution of "a real" FRW universe. But
we do think that the age-problem of the universe (the problem of the large
Hubble constant) is non-existent.

\appendix
\section{The Zeta-Function for Spin 1/2 Fermions
       Minimally Coupled to a Gauge Field}
For real world purposes one should consider the case where the fermion field couples
to a gauge boson field.  The gauge invariant and
 space covariant derivative is
\beq
    D_m=e^\mu_m(\bd_\mu+\frac{i}{2}\omega^{pq}_\mu(x)X_{pq}+igA^a_\mu(x)T_a)
\eeq
Using the representation (??) and the relations
\beqa
    \gamma^\mu\gamma^\nu&=&\eta^{pq}+2i\sigma^{pq}\\
    \left[\gamma^m,\sigma^{pq}\right]&=&i(\eta^{mp}\gamma^q-\eta^{mq}\gamma^p)\\
    \left\{\gamma^p\sigma_{mn},\gamma^q\right\}&=&2\eta^{pq}\sigma_{mn}+
    \delta^q_m\delta^p_n-2\delta^q_m\sigma^p_{~n}                  
                            -i\delta^q_n\delta^p_m+2\delta^q_n\sigma^p_{~m}
\eeqa 
one obtains, for the derivative squared            
\beqa  
D^2\hspace{-5mm}/&=&(\Dslash\bd+\frac{i}{2}e^\mu_m\gamma^m\omega_\mu^{pq}\sigma_{pq}+g\gamma^mA_m^aT_a)
 (\Dslash\bd+\frac{i}{2}e^\mu_m\gamma^m\omega_\mu^{pq}\sigma_{pq}+g\gamma^mA_m^aT_a)\nonumber\\
&=&\Box + \xi_fR+g^2\gamma^p\gamma^{p'}A^a_pA^b_{p'}T_aT_b + g\gamma^q\gamma^p(\bd_qA_p^a)T_a
 \nonumber\\
    &&\hspace{10mm}+ \frac{i}{2}g\gamma^m\sigma_{pq}\gamma^{p'}\omega_\mu^{pq}e^\mu_mA^b_{p'}T_b
 + \frac{i}{2}g\gamma^p\gamma^{m'}\sigma_{p'q'}A^a_pT_a\omega^{p'q'}_\nu e^\nu_{p'}
\nonumber\\
&=&\Box +\xi_fR+2g\sigma^{pq}F^a_{pq}T_a+g^2\eta^{pq}A^a_pA^b_qT_aT_b+g(\bd^pA_p^a)T_a\nonumber\\
      &&\hspace{10mm}+\frac{i}{2}gT_a\left(\sigma_{pq}\omega^{pq}_\mu e_x^xA^a_x+i\omega^{mn}_\mu 
    (e^\mu_mA^a_n-e^\mu_nA^a_m)-4\omega^{nq}_\mu e^\mu_mA^a_n\sigma^m_{~q}\right)\nonumber\\
&\equiv&\Box +\xi_fR+2g\sigma^{pq}F^a_{pq}T_a+g^2\eta^{pq}A^a_pA^b_qT_aT_b+{\cal G}(A)
\eeqa
To make this expression manageable note that putting
\beq
{\cal G}(A)=0
\eeq
is an allowed gauge condition. To show this  one needs to demonstrate that $\det(\frac{\delta{\cal G}}{\delta\omega^a})\neq 0$
which is easily done. Perform an infinitesimal gauge transformation $\delta A_m^a=D_m\omega^a$ (where $\omega^a$ is the
arbitrary function appearing in the transformation rule for fermions $\psi\rightarrow \exp(ig\omega^aT_a)\psi$ in $\cal G$ to get
\beqs
    \frac{\delta {\cal G}}{\delta \omega^a} \equiv\eta^{mn}\bd_mD^n+\mbox{function}
\eeqs
Since Lorentz gauge is an allowed gauge condition, we know that $\det \bd_mD^m\neq 0$, and adding a function cannot change this
fact. Hence ${\cal G}(A)=0$ is good gauge condition.\\
Using the mean field approximation described in appendix A, the gauge field dependent terms simply becomes a function which is
in principle no different from the $\xi_fR$ term to which it, for calculational purposes, can be added.

\section{Calculation of $\langle x_ix_j\rangle$}
We want to derive an expression for $\langle x_ix_j\rangle$ where $x_i,x_j$
are components of an $N$-dimensional vector, and where
\beqs
	\langle x_ix_j\rangle =\frac{\int x_ix_je^{-x^tMx}d^Nx}{\int e^{-x^tMx}
	d^Nx}
\eeqs
with $M$ some symmetrical $N\times N$-matrix.\\
First note that the integral vanishes whenever $i\neq j$. Secondly, that
since $M$ is symmetrical we can diagonalise it in which case the problem
reduces to the $N=1$ case.\\
For $N=1$ we have
\beq
	\langle x^2\rangle = \frac{1}{2}M^{-1}
\eeq
For arbitrary $N$ we then have
\beq
	\langle x_ix_j\rangle = \frac{1}{2}\delta_{ij} (M^{-1})_{ii}
\eeq
where {\em no} summation over $i$ is to be performed. This is the needed
result.

%\end{document}
\newpage
\begin{figure}
\caption{Evolution of the scale-factor for: a) $K=+1$,  b) $K=0$,  c) $K=-1$
with a minimally coupled scalar field.}
\end{figure}
\begin{figure}
\caption{Evolution of the scale factor for:  a) $K=+1$,  b) $K=0$  and  
c) $K=-1$ with a conformally coupled scalar field}
\end{figure}

\begin{figure}
\caption{Evolution of scale factor for the cases of the residing field 
         being a minimally coupled (solid line) or a conformally coupled
         (dotted line):  a) $K=+1$,  b) $K=-1$.}
\end{figure}

\begin{figure}
\caption{Evolution of the scale factor for : a) $K=+1$, b) $K=0$, c) $K=-1$
with a fermion field.}
\end{figure}

\end{document}